\documentclass[conference]{IEEEtran}
\usepackage[pdftex]{graphicx}
\usepackage{tabularx}
\usepackage{amsmath} 
\usepackage{multirow,booktabs}
\usepackage{multicol}
\usepackage[justification=centering]{caption}
\usepackage{mathabx}
\usepackage{changepage}
\usepackage{algorithm}

\usepackage{enumitem}
\usepackage{url}
\usepackage{hyperref} 

\usepackage{cite}

\begin{document}

%
\title{{Boosting in Image Quality Assessment}}

\author{
\IEEEauthorblockN{Dogancan Temel}
\IEEEauthorblockA{Center for Signal and \\Information Processing\\School of Electrical and\\Computer Engineering\\
	Georgia Institute of Technology\\
	Atlanta, Georgia 30332--0250\\
	Email: dcantemel@gmail.com}
\and
\IEEEauthorblockN{Ghassan AlRegib}
\IEEEauthorblockA{Center for Signal and \\Information Processing\\School of Electrical and\\Computer Engineering\\
	Georgia Institute of Technology\\
	Atlanta, Georgia 30332--0250\\
	Email: alregib@gatech.edu}
}

%

\onecolumn 

\begin{description}[labelindent=1cm,leftmargin=3cm,style=multiline]

\item[\textbf{Citation}]{D. Temel and G. AlRegib, "Boosting in image quality assessment," 2016 IEEE 18th International Workshop on Multimedia Signal Processing (MMSP), Montreal, QC, 2016, pp. 1-6.
} \\

\item[\textbf{DOI}]{\url{https://doi.org/10.1109/MMSP.2016.7813335}} \\

\item[\textbf{Review}]{Date added to IEEE Xplore: 16 January 2017} \\


\item[\textbf{Bib}] {
@INPROCEEDINGS\{Temel2016\_MMSP,\\ 
author=\{D. Temel and G. AlRegib\}, \\
booktitle=\{2016 IEEE 18th International Workshop on Multimedia Signal Processing (MMSP)\},\\ 
title=\{Boosting in image quality assessment\},\\ 
year=\{2016\},\\
pages=\{1-6\},\\ 
doi=\{10.1109/MMSP.2016.7813335\},\\ 
ISSN=\{2473-3628\},\\ 
month=\{Sept\},\} 
} \\

\item[\textbf{Copyright}]{\textcopyright 2016 IEEE. Personal use of this material is permitted. Permission from IEEE must be obtained for all other uses, in any current or future media, including reprinting/republishing this material for advertising or promotional purposes,
creating new collective works, for resale or redistribution to servers or lists, or reuse of any copyrighted component
of this work in other works. } \\

\item[\textbf{Contact}]{\href{mailto:alregib@gatech.edu}{alregib@gatech.edu}~~~~~~~\url{https://ghassanalregib.com/}\\ \href{mailto:dcantemel@gmail.com}{dcantemel@gmail.com}~~~~~~~\url{http://cantemel.com/}}
\end{description} 

\thispagestyle{empty}
\newpage
\clearpage

\twocolumn

\maketitle
\IEEEpeerreviewmaketitle

\begin{abstract}
In this paper, we analyze the effect of boosting in image quality assessment through multi-method fusion. Existing multi-method studies focus on proposing a single quality estimator. On the contrary, we investigate the generalizability of multi-method fusion as a framework. In addition to support vector machines that are commonly used in the multi-method fusion, we propose using neural networks in the boosting. To span different types of image quality assessment algorithms, we use quality estimators based on fidelity, perceptually-extended fidelity, structural similarity, spectral similarity, color, and learning. In the experiments, we perform k-fold cross validation using the LIVE, the multiply distorted LIVE, and the TID 2013 databases and the performance of image quality assessment algorithms are measured via  accuracy-, linearity-, and ranking-based metrics. Based on the experiments, we show that boosting methods generally improve the performance of image quality assessment and the level of improvement depends on the type of the boosting algorithm. Our experimental results also indicate that boosting the worst performing quality estimator with two or more additional methods leads to statistically significant performance enhancements independent of the boosting technique and neural network-based boosting outperforms support vector machine-based boosting when two or more methods are fused. 
\end{abstract}

\section{Introduction}
\label{sec:intro}
\IEEEPARstart{I}mages have dominated online media and social networks because of the advances in capturing, storage, streaming, and display technologies. Everyday, on average, billions of images are shared online. To be able to share all these images within a limited bandwidth, we need to design compact representations that not only satisfy bandwidth requirements but also maintain perceived quality. The ever-increasing number of images makes it impossible to assess the perceived quality of all publicly available images subjectively. Therefore, there is an imminent need to automatically assess the quality of experience (QoE), whose definition depends on the application and in case of imaging applications, the core of QoE is image quality.

In the research community, image quality assessment is modeled by mapping images to subjective scores. Quality estimators are designed to process an image or images to provide a quality score. Fidelity-based approaches measure pixel-wise differences and they can be extended with visual system-based characteristics to obtain a more perceptually correlated quality score \cite{Egiazarian, Ponomarenko2007, Ponomarenko2011 }. Instead of quantifying pixel-wise differences,  measuring structural similarity is also commonly used in the literature \cite{Wang2003, Inter2005, Wang2011}. Even the majority of the quality estimators use only grayscale images or intensity channels, color information is also used in the literature \cite{Carnec2008, Zhang2011,Temel2015}. In addition to these hand-crafted quality estimators, methods that are based on modeling natural scene statistics and data-driven learning are also used in the literature \cite{Sheikh2005, LeCallet2006, Kang, UNIQUE}. The majority of the quality estimators refer to visual system characteristics but none of them is a comprehensive model of the perception process. Existing image quality estimators differ from each other in various ways. However, all these methods fundamentally map pixels to  subjective scores. Moreover, even some of the methods are less perceptually correlated than others, they can still contain additional information that can not be provided by better performing methods. Therefore, multiple methods can be fused to boost the overall performance. Boosting is initially discussed in \cite{kearns1988} and \cite{kearns1994 } to investigate whether it is possible to obtain strong learners from weak learners or not. In \cite{schapire1990}, the authors describe a method for converting a weak learning algorithm into a strong one that obtains arbitrarily high accuracy.

Based on the boosting discussion, we can also convert image quality assessment algorithms with poor performance into highly perceptually correlated quality estimators. In \cite{leontaris2007}, the authors analyze the performance of multiple methods and combine two methods linearly to obtain a hybrid quality estimator. Linear weights are selected by exploring a parameter space. The authors in \cite{liu2011} propose a regression-based approach that is used to fuse quality estimates of multiple methods non-linearly. In addition to estimating a quality score directly, hand-crafted features are also used to classify distortion types and a regression approach is used in each distortion type separately to learn the mapping function. In \cite{liu2013}, the multi-method fusion is extended with a method selection algorithm to reduce the overall complexity. The authors in \cite{liu2015} follow a regression-based approach to obtain two types of image quality estimators that are separately trained with features of existing quality estimators and hand-crafted features that measure degradations overlooked by the existing features. The scores of individual quality estimators are fused with a support vector regression stage along with a statistical testing-based selection mechanism. A similar parallel boosting approach based on support vector regression is also used for stereoscopic image quality assessment in \cite{ko2016}.

Multi-method fusion approaches are promising if they are considered as a framework, which can lead to more comprehensive quality estimators as the boosting methods and the fused image quality assessment algorithms improve. The framework should not be limited to some specific quality estimators, distortion types, or learning methods. To investigate the generalizability of fusing quality estimators, we use $11$ image quality assessment algorithms that can be grouped into $5$ different categories as fidelity-based, perceptually-extended fidelity-based, structural similarity-based, color-based, and learning-based. In addition to performing boosting with support vector machines, we also use neural networks. Fused methods are trained and tested over three different databases that include distortion types based on compression, image noise, communication, blur, color, global, and local. Finally, performances of fused methods on different databases are measured using linearity-, accuracy-, and ranking-based metrics. At first, we investigate the performance of regressed versions of the existing methods and compare the performance with existing methods. Then, we analyze the performance of multi-method fusion that uses all the quality estimators. Finally, performance of boosted methods with respect to the number of fused methods are analyzed. To the best of our knowledge, there are only a few studies that fuse multiple methods, which use a single type of architecture for regression. In this work, in addition to commonly used support vector machines, we propose using neural networks in the multi-method fusion. In Section \ref{setup}, we describe the experimental setup, which includes brief descriptions of used image quality estimators, boosting methods, databases, data partitioning, number of experiments, and performance metrics. We discuss the experimental results in Section \ref{results} and conclude our discussion in Section \ref{sec:conc}.


\vspace{4.0mm}
\section{Experimental Setup}
\label{setup}

\subsection{Image Quality Estimators}
\label{iqa}

\subsubsection{Fidelity-based}
\label{iqa_fidelity}
Fidelity attributes quantify the changes in a degraded image with respect to a reference image and they are commonly preferred in image and video coding standards for rate-distortion optimization because of low computational complexity and ease of implementation. The intuitive method to measure the fidelity of an image is to directly compare it with its distortion-free image, if available. Mean square error (MSE) is a commonly used pixel-wise fidelity method, which is calculated by obtaining the difference between images, taking the square root of the difference, and calculating the mean value. MSE is scaled by the range of an image and mapped with a logarithmic function to obtain the peak signal-to-noise ratio (PSNR), which is one of the quality estimators used in the boosting operations.
\subsubsection{Perceptually Extended Fidelity-based}
\label{iqa_structure}
Image quality metrics use the characteristics known about the visual system to make the perceptual quality assessment more accurate. The authors in \cite{Egiazarian} extend PSNR by removing mean shift, stretching contrast block-wise, and quantizing DCT coefficients with the compression table proposed by JPEG. These extensions are performed to make PSNR compatible with the human visual system and the extended metric is named as PSNR-HVS. Reduction by value of contrast masking is also added to the metric and the modified version is named as PSNR-HVS-M \cite{Ponomarenko2007}.  These metrics are further extended by adding contrast change and mean shifting sensitivity (PSNR-HA, PSNR-HMA) as explained in \cite{Ponomarenko2011}, both of which are used in the boosting operations.

\subsubsection{Structural Similarity-based}
\label{iqa_structure}
Structural similarity is commonly obtained by quantifying the similarity between mean subtracted and divisive normalized images. The authors in \cite{Wang2003} propose a full reference metric (SSIM) based on the comparison between a reference and a distorted image in terms of luminance, contrast, and structure in the spatial domain. These structure-based methods are also extended to multi-scale (MS-SSIM) \cite{Wang2003}, complex domain (CW-SSIM) \cite{Inter2005}, and  information-weighted (IW-SSIM) \cite{Wang2011} versions. All of these structural similarity methods are used in the boosting operations. Moreover, we also use spectral similarity in the boosting \cite{Zhang2012}.

\subsubsection{Color-based}
\label{iqa_structure}
The human visual system (HVS) is more sensitive to changes in intensity compared to color  \cite{Lambrecht2001}.  Although color may not be as informative as intensity, it can still contain additional information.  An intuitive way to use color information in the image quality assessment is pixel-wise fidelity. FSIMc \cite{Zhang2011}  and PerSIM \cite{Temel2015} introduce color information by computing pixel-wise fidelity over chroma channels in the La*b* color space. In addition to the color-based similarity,  FSIMc computes similarity based on phase congruency and gradient magnitude, and PerSIM computes similarity based on band-pass features that are obtained from the contrast sensitivity formulation of the retinal ganglion cells. FSIMc are PerSIM are used in the boosting operations.

\subsubsection{Learning-based}
\label{iqa_structure}
It is not possible to handcraft a comprehensive quality estimator that covers all the aspects of visual system. Therefore, data-driven approaches can be used to design quality estimators. The majority of the data-driven approaches require distortion-specific images or subjective scores in the training, which can bias the performance of boosting methods. 	Therefore, we use the data-driven quality estimator UNIQUE, which is trained with solely generic images in an unsupervised fashion. Images are pre-processed with a mean subtraction stage, a whitening operation, and color space transformations to obtain more descriptive representations in terms of structure and color. These representations are fed to a linear decoder to obtain sparse representations. An objective score is obtained by comparing the sparse representations in terms of monotonic behavior.

\subsection{Boosting Methods}
\label{boosting}
Rather than using specifically tuned deep networks or complicated architectures, we analyze the effect of boosting through two off-the-shelf methods. We use a generic neural network and a support vector machine. The only parameter that we adjust in the neural network architecture is the number of neurons in a single hidden layer, which is set to the total number of quality estimators used in the experiments. By default, we use mean square error as the cost function and Levenberg-Marquardt as the training function, which does not necessarily guarantee a global minimum. The default configuration in a support vector machine includes a sequential minimal optimization (SMO) as the solver and a linear kernel.

\subsection{Databases}
In the performance comparison of the quality estimators, we use
the LIVE \cite{live2006}, the multiply distorted LIVE (MULTI) \cite{multi2012}, and the TID 2013 (TID13) databases \cite{tid13journal}. The distortion types in these databases can be grouped into seven categories as given in Table \ref{tab_db}. JPEG, JPEG2000, and lossy compression of noisy images are included in the compression category. The noise category consists of Gaussian noise, additive noise in color components which is more intensive than additive noise in the luminance component, spatially correlated noise, masked noise, high frequency noise, impulse noise, quantization noise, image denoising, multiplicative Gaussian noise, comfort noise, and lossy compression of noisy images. JPEG and JPEG2000 transmission errors are included in the communication category, and  Gaussian blur and sparse sampling and reconstruction are in the blur category.  The color category consists of change of color saturation, image color quantization with dither and chromatic aberrations. Intensity shift and contrast change are included in the global category, and non-eccentricity pattern noise and local block-wise distortions are in the local category.

\begin{table}[htbp!]
\centering
\caption{The number of distorted images in each database.}
\label{tab_db}
\begin{tabular}{c|cccc}
\hline
                    & {\bf LIVE \cite{live2006}} & {\bf MULTI \cite{multi2012}} & {\bf TID13 \cite{tid13journal}} & {\bf Total} \\ \hline
{\bf Compression}   & 460        & 180         & 375         & 1015        \\ 
{\bf Noise}         & 174        & 180         & 1375        & 1729        \\ 
{\bf Communication} & 174        & -           & 250         & 424         \\ 
{\bf Blur}          & 174        & 315         & 250         & 739         \\ 
{\bf Color}         & -          & -           & 375         & 375         \\ 
{\bf Global}        & -          & -           & 250         & 250         \\ 
{\bf Local}         & -          & -           & 250         &250          \\ \hline
\end{tabular}
\end{table}


\subsection{Data Partition and Number of Experiments}
In the experiments, the performance of the quality estimators are measured with k-fold cross validation, in which $k$ is set to $5$. At each iteration,  $20\%$ of total images in each database are selected as the test set. In Section \ref{results_part2}, we test the performance of methods boosted with a neural network and a support vector machine. Each method is trained and tested $100$ times. The test set in each iteration is also used to measure the performance of existing quality estimators. Since there are $11$ different quality estimators, $2$ boosting methods, and $100$ runs, we report the average performance of existing quality estimators for $2,200$ runs in Section \ref{results_part1}.

\subsection{Performance Metrics}
We use accuracy-, linearity-, ranking-, and statistical significance-based metrics in the performance analysis and comparison. Before the metric calculations, a mapping operation is performed between objective and subjective scores as suggested in \cite{Wang2006b}. The mapping formulation used in the simulations can be expressed as
\begin{equation}
\label{eq:nonlinreg}
V=\beta_1 \left ( \frac{1}{1}-\frac{1}{2+exp(\beta_2(V_0 -\beta_3 ))} \right )+\beta_4 V_0 +\beta_5,
\end{equation}
where $V_0$ is an estimated score, $V$ is a regressed output and $\beta$s are tuning parameters that are set according to the relationship between the quality estimates and the subjective scores.
\vspace{2.0mm}
\subsubsection{Accuracy}
Root mean square error measures the accuracy of the quality estimators as 
\begin{equation}
RMSE=\sqrt{\frac{\sum_{s=1}^{N}(x_s-y_s)^2}{N}},
\end{equation}
where $x_s$ is an estimated score and $y_s$ is a subjective score corresponding to an image indexed with $s$, and $N$ is the total number of images.
\vspace{2.0mm}

\subsubsection{Linearity}
 Pearson correlation coefficient is used to measure the linearity of the predictions which is formulated as
\begin{equation}
\label{eq:pearson}
PLCC=\frac{\sum_{s=1}^{N}(x_s-\mu_x)(y_s-\mu_y)}{\sqrt{\sum_{s=1}^{N}(x_s-\mu_x)^2}\cdot \sqrt{\sum_{s=1}^{N}(y_s-\mu_y)^2}},
\end{equation}
where $x_s$ is an estimated score and $y_s$ is a subjective score corresponding to an image indexed with $s$,  $\mu$ is the average operator, and $N$ is the total number of images. 
\vspace{2.0mm}

\subsubsection{Ranking}
Spearman correlation is used to measure the monotonic relationship between quality estimates and subjective scores. Instead of using the exact values, rank of the values are used. The formulation of Spearman correlation coefficient is given as
\begin{equation}
\label{eq:spearman}
SRCC=1-\frac{6\sum_{s=1}^{N}(X_s-Y_s)^2}{N\cdot (N^2-1)},
\end{equation}
where $X_s$ is a rank assigned to a score $x_s$ and $Y_s$ is a rank assigned to a subjective score $y_s$, which correspond to an image indexed with $s$, and $N$ is the total number of images.
\vspace{2.0mm}

\subsubsection{Statistical Significance}
\label{subsubsec_stats_significance}
In order to assess the significance of the difference between correlation coefficients, we use the statistical significance tests suggested in ITU-T Rec. P.1401 \cite{ITU2012}.

\begin{center}
\begin{table*}[htbp!]
\scriptsize
\centering
\caption{Performance of existing IQA methods using 5-fold validation for 2,200 runs. }
\label{tab_part1}
\begin{tabular}{c|ccccccccccc}
\hline
               & \textbf{PSNR} & \textbf{PSNR-HA} & \textbf{PSNR-HMA} & \textbf{SSIM} & \textbf{MS-SSIM} & \textbf{CW-SSIM} & \textbf{IW-SSIM} & \textbf{SR-SIM}  & \textbf{FSIMc} & \textbf{PerSIM} &\bf UNIQUE    \\ \cline{2-12}
         
               & \multicolumn{11}{c}{\textbf{Root Mean Square Error}}                                                                                                                                                                        \\ \hline
\textbf{LIVE}      &  8.60  &  6.92 &  \bf 6.57 &   7.51  &  7.42  & 11.3  &  7.09   & 7.53  &  7.19  &  6.79  &  6.75
 \\ 
\textbf{MULTI}    & 12.7  & 11.2  & 10.7  & 11.0  & 11.2  & 18.8 &  10.0 & \bf  8.68 &  10.7  &  9.89 &   9.24
  \\ 
\bf TID13      & 0.87  &  0.65  &  0.69  &  0.76  &  0.69  &  1.20 &   0.68 &   0.61  &  0.68   & 0.64 & \bf  0.61
 \\ \hline                  
               & \multicolumn{11}{c}{\textbf{Pearson Correlation Coefficicent}}                                                                                                                                                                        \\ \hline
\textbf{LIVE}      &  0.927  &  0.953 & \bf  0.958  &  0.945 &   0.947 &  0.871 &   0.951 &   0.945 &   0.950 &   0.955 &   0.956  \\ 
\textbf{MULTI}        &  0.737   & 0.799 &   0.819  &  0.813  &  0.803  &  0.406 &   0.846  & \bf 0.887 &   0.820 &   0.850  &  0.871    \\ 
\bf TID13      & 0.705  &  0.850 &   0.827   & 0.788 &   0.830  &  0.228  &  0.831 &   0.866  &  0.832 &   0.854 &  \bf 0.868
\\ \hline        
\textbf{}      & \multicolumn{11}{c}{\textbf{Spearman Correlation Coefficient}}                                                                                                                                                                        \\ \hline
\textbf{LIVE}      &  0.907  &  0.936 &   0.942  &  0.947  &  0.949  &  0.900 & \bf  0.959  &  0.954  &  0.958   & 0.949  &  0.950    \\ 
\textbf{MULTI}      &  0.672  &  0.709  &  0.738  &  0.855 &   0.831  &  0.626 & \bf  0.878  &  0.860  &  0.860 &   0.812 &   0.861  \\ 
\bf TID13    &  0.700  &  0.846  &  0.816  &  0.740 &   0.784  &  0.562 &   0.776 &   0.806  &  0.850  &  0.852  & \bf 0.859
\\ \hline

\end{tabular}
\vspace{3.00mm}
\end{table*}
\end{center}

\section{Experimental Results}
\label{results}

\subsection{Part 1}
\label{results_part1}
We report the performance of existing quality estimators in Table \ref{tab_part1}. 
In terms of root mean square error and Pearson correlation, the best performing methods are PSNR-HMA in the LIVE database and SR-SIM in the MULTI database. In terms of Spearman correlation, IW-SSIM is the best performing method in the LIVE and the MULTI databases. UNIQUE is the best performing quality estimator in terms of all the metrics in the TID13 database. 

\begin{center}
\begin{table*}[htbp!]
\footnotesize
\centering
\caption{Performance of IQA methods with neural network-based regression using 5-fold validation for 100 runs.}
\label{tab_part1_nn}
\begin{tabular}{l|lllllllllll}
\hline
               & \textbf{PSNR} & \textbf{PSNR-HA} & \textbf{PSNR-HMA} & \textbf{SSIM} & \textbf{MS-SSIM} & \textbf{CW-SSIM} & \textbf{IW-SSIM} & \textbf{SR-SIM}  & \textbf{FSIMc} & \textbf{PerSIM} &\bf UNIQUE     \\ \cline{2-12}
         
               & \multicolumn{11}{c}{\textbf{Root Mean Square Error}}                                                                                                                                                                        \\ \hline
\textbf{LIVE}  &  8.21  &  6.93  &  6.61 &   5.98  &  5.91  &  8.81   & \bf 5.49  &  5.80  &  5.56  &  6.35 &   6.09 
   \\ 
\textbf{MULTI}  & 13.3 &  11.4 &  10.8 &   8.68  &  9.56 &  14.6  & \bf 7.91 &   8.29  &  8.12  &  9.76  &  8.80        
\\ 
\bf TID13    &  0.86  &  0.64  &  0.68 &   0.72  &  0.64 &   1.70 &   0.68 &   \bf 0.59 &   0.60  &  0.62  &  0.61  
\\ \hline

               & \multicolumn{11}{c}{\textbf{Pearson Correlation Coefficicent}}                                                                                                                                                                        \\ \hline
\textbf{LIVE} &0.934   & 0.954   & 0.957   & 0.966   & 0.967    &0.923   & \bf 0.971  &  0.967  &  0.970  &  0.961 &   0.964  
  \\ 
\textbf{MULTI}     & 0.710 &   0.793 &   0.821  &  0.890 &   0.866 &   0.646 &  \bf 0.907  &  0.900  &  0.903  &  0.855  &  0.886 \\ 
\bf TID13  & 0.722 &   0.852  &  0.830   & 0.814 &   0.852 &   0.442 &   0.836 &  \bf 0.879  &  0.872  &  0.864  &  0.870   
 \\ \hline        

\textbf{}      & \multicolumn{11}{c}{\textbf{Spearman Correlation Coefficient}}                                                                                                                                                                        \\ \hline
\textbf{LIVE} & 0.904  &  0.937  &  0.941   & 0.947  &  0.950   & 0.894 & \bf  0.958 &   0.954  &  0.957  &  0.948  &  0.950    
  \\ 
\textbf{MULTI} & 0.660  &  0.701  &  0.738  &  0.858 &   0.827   & 0.613 & \bf  0.883  &  0.874 &   0.872  &  0.797  &  0.861     \\ 
\bf TID13    & 0.706 &   0.845  &  0.813   & 0.794  &  0.834  &  0.558  &  0.807  &  0.843 &   0.850  &  0.852  & \bf 0.860     \\ \hline        

\end{tabular}
\vspace{3.00mm}

\end{table*}
\end{center}

\begin{center}

\begin{table*}[htbp!]
\footnotesize
\centering
\caption{Performance of IQA methods with support vector machine-based regression using 5-fold validation for 100 runs.}
\label{tab_part1_svm}
\begin{tabular}{c|ccccccccccc}
\hline
               & \textbf{PSNR} & \textbf{PSNR-HA} & \textbf{PSNR-HMA} & \textbf{SSIM} & \textbf{MS-SSIM} & \textbf{CW-SSIM} & \textbf{IW-SSIM} & \textbf{SR-SIM}  & \textbf{FSIMc} & \textbf{PerSIM} &\bf UNIQUE     \\ \cline{2-12}
         
               & \multicolumn{11}{c}{\textbf{Root Mean Square Error}}                                                                                                                                                                        \\ \hline
\textbf{LIVE}       & 8.58  &  7.00 & \bf  6.66 &   7.58 &   7.51  & 11.90  &  7.10   & 7.76  &  7.27  &  6.89 &   6.79    
   \\ 
  \textbf{MULTI}  & 13.1 &  11.3  & 10.7  & 11.0  & 11.3  &18.5  & 10.0    & \bf 8.75  & 10.9 &  10.0  &  9.23 
\\ 
\bf TID13 & 0.88  &  0.66   & 0.70   & 0.77  & 0.69  &  1.21   & 0.69   & 0.62 &   0.69  &  0.64  & \bf 0.61   \\ \hline        
        
               & \multicolumn{11}{c}{\textbf{Pearson Correlation Coefficient}}                                                                                                                                                                        \\ \hline
\textbf{LIVE }    &  0.928   & 0.953   & \bf 0.958  &  0.945   & 0.947   & 0.871   & 0.952   & 0.945   & 0.951   & 0.955   & 0.956  
  \\ 
\textbf{MULTI}    & 0.720   & 0.798 &   0.820  &  0.814  &  0.798  &  0.399 &   0.851 &  \bf 0.887 &   0.816  &  0.847  &  0.874  \\ 
\bf TID13    & 0.706  &  0.848  &  0.826  &  0.787  &  0.828  &  0.226 &   0.829 &   0.866 &   0.833  &  0.855 &  \bf 0.869    
\\ \hline        

\textbf{}      & \multicolumn{11}{c}{\textbf{Spearman Correlation Coefficient}}                                                                                                                                                                        \\ \hline
\textbf{LIVE}  & 0.908   & 0.935   & 0.942  &  0.947    &0.950    &0.901    &\bf 0.960    &0.953   & 0.958   & 0.948  &  0.951 
  \\ 
\textbf{MULTI}  & 0.652    &0.714   & 0.738   & 0.855  &  0.828   & 0.616   & \bf 0.882    &0.860   & 0.855   & 0.813   & 0.864
 \\ 
\bf TID13  & 0.701   & 0.843 &   0.814  &  0.736 &   0.785 &   0.558 &   0.773 &   0.807  &  0.849  &  0.853 &  \bf 0.859   \\ \hline        
\end{tabular}
\vspace{3.00mm}

\end{table*}
\end{center}
\vspace{-12.0mm}
Neural network-based regression results are given in Table \ref{tab_part1_nn}.
 Neural networks trained with fidelity-, perceptually-extended fidelity-, and perceptual similarity-based methods enhance the performances in some categories and degrade in others with minor changes. In terms of root mean square error and Pearson correlation, neural networks lead to significant or minor enhancements for structural, spectral, unsupervised learning-based, and feature-based similarity methods. In terms of Spearman correlation, neural networks mostly lead to some minor changes other than some major changes in the TID13 database. After the neural network-based regression, IW-SSIM becomes the best performing quality estimator in terms of root mean square error and Pearson correlation in the LIVE and the MULTI databases. In the TID13 database, SR-SIM becomes the best performing quality estimator in terms of root mean square error and Pearson correlation after the neural network-based regression. We also perform support vector machine-based regression and the results are given in Table \ref{tab_part1_svm}. Support vector regression does not lead to significant changes when it is only trained with one method and the best performing methods are the same with the existing methods. In Table \ref{tab_best}, we report the best performance values of existing and regressed methods. Moreover, we also report the performances of neural network- and support vector machine-based boosting. Existing methods regressed with neural networks perform better than existing methods in all the categories other than Spearman in the LIVE database, and the performances of existing methods regressed with support vector machines are similar to existing methods. Support vector machine-based boosting performs better than existing and regressed existing methods in the MULTI and the TID13 databases whereas in the LIVE database, it is better in some categories and worse in others. Neural network-based boosting leads to the best performances in all the categories.

\begin{center}
\begin{table}[htbp!]
\footnotesize
\centering
\caption{Performance of existing, regressed, and boosted IQA methods.}
\label{tab_best}
\begin{tabular}{c|ccccc}
\hline
                &\bf Existing   & \bf NN &\bf SVR &\bf NN&\bf SVR    \\ 
                &\bf Best&\bf Best &\bf Best&\bf Boost&\bf Boost \\ \hline
         
               & \multicolumn{5}{c}{\textbf{Root Mean Square Error}}                                                                                                                                                                        \\ \hline

\textbf{LIVE}   &6.57&5.49&6.66&4.54&5.62    \\ 
\textbf{MULTI}  &8.68&7.91&8.75&6.73&7.07     \\ 
\bf TID13 &0.61&0.59&0.61&0.45&0.51 \\ \hline            
         
               & \multicolumn{5}{c}{\textbf{Pearson Correlation Coefficicent}}                                                                                                                                                                        \\ \hline
\textbf{LIVE}    &0.958&0.971&0.958&0.980&0.970 \\
\textbf{MULTI}   &0.887&0.907&0.887&0.934&0.926\\ 
\bf TID13 &0.868&0.879&0.869&0.931&0.909 \\ \hline
\textbf{}      & \multicolumn{5}{c}{\textbf{Spearman Correlation Coefficient}}                                                                                                                                                                        \\ \hline
\textbf{LIVE} &0.959&0.958&0.960&0.969&0.956\\ 
\textbf{MULTI} &0.878&0.883&0.882&0.918&0.915 \\ 
\bf TID13 &0.859&0.860&0.859&0.921&0.895 \\ \hline
\end{tabular}
\end{table}
\end{center}

\subsection{Part 2}
\label{results_part2}
\begin{center}

\begin{figure*}[htbp!]
\begin{minipage}[b]{0.32\linewidth}
  \centering
\includegraphics[width=\linewidth]{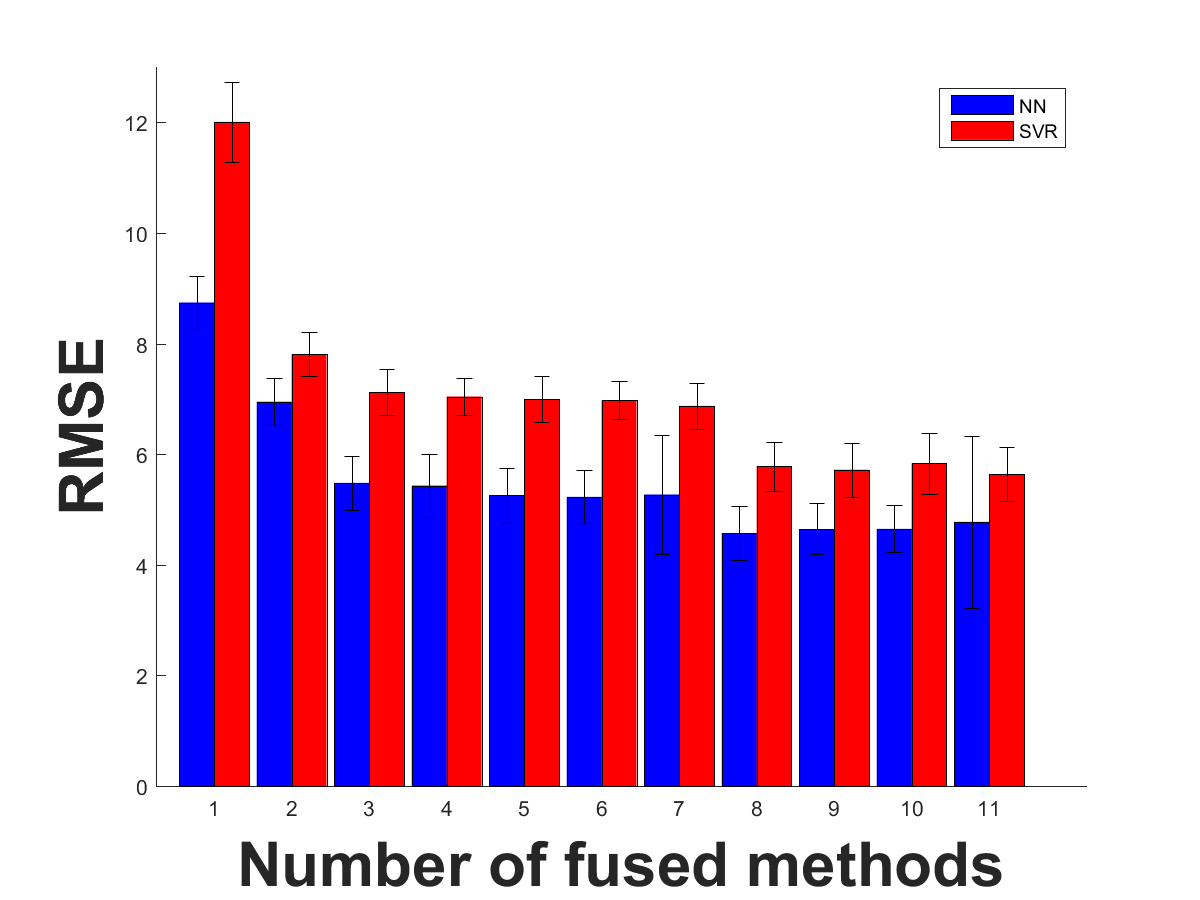}
  \vspace{0.03cm}
  \centerline{\footnotesize{(a)LIVE-RMSE}}

\end{minipage}
\hfill
\begin{minipage}[b]{0.32\linewidth}
  \centering
\includegraphics[width=\linewidth]{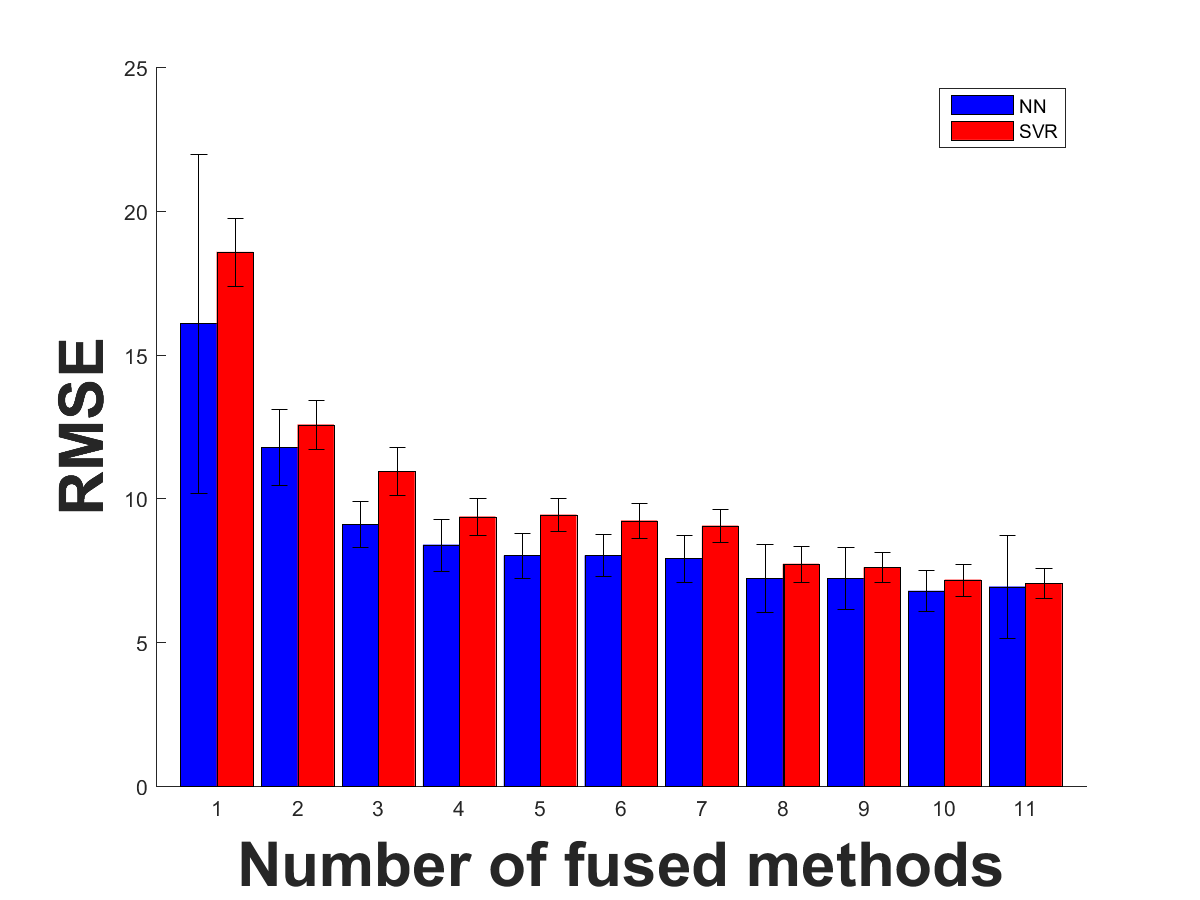}
  \vspace{0.03cm}
  \centerline{\footnotesize{(b) MULTI-RMSE }}

\end{minipage}
\hfill
\begin{minipage}[b]{0.32\linewidth}
  \centering
\includegraphics[width=\linewidth]{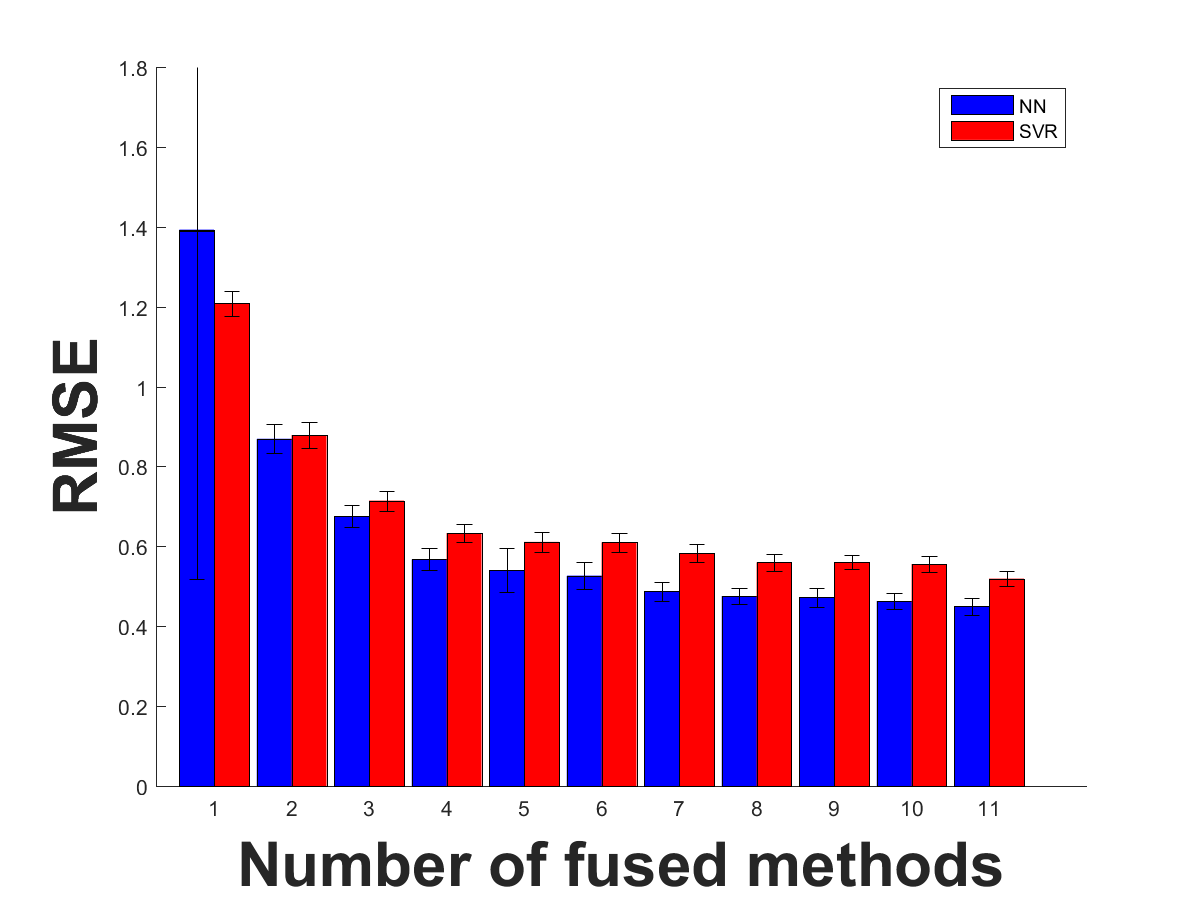}
  \vspace{0.03 cm}
  \centerline{\footnotesize{(c) TID-RMSE   } }
\end{minipage}

\begin{minipage}[b]{0.32\linewidth}
  \centering
\includegraphics[width=\linewidth]{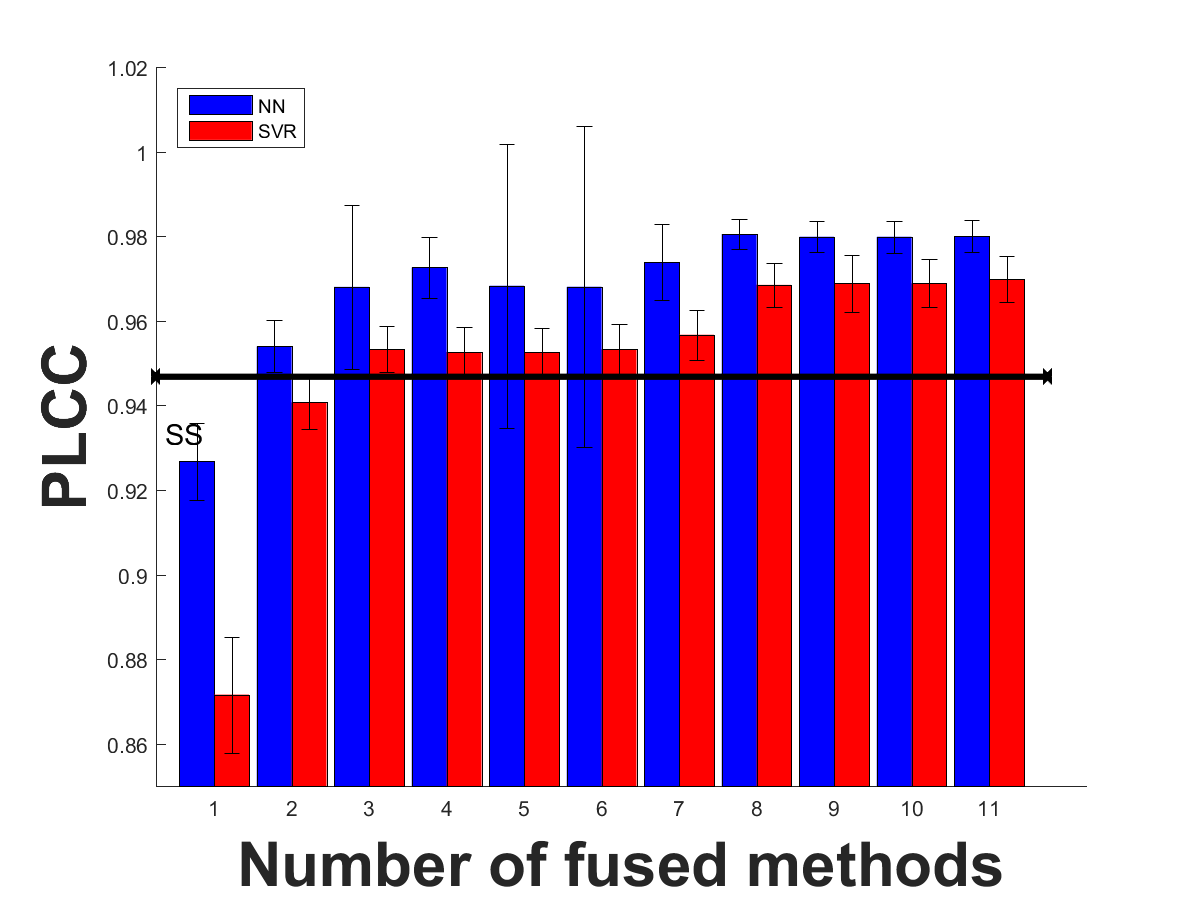}
  \vspace{0.03cm}
  \centerline{\footnotesize{(d)LIVE-PLCC}}

\end{minipage}
\hfill
\begin{minipage}[b]{0.32\linewidth}
  \centering
\includegraphics[width=\linewidth]{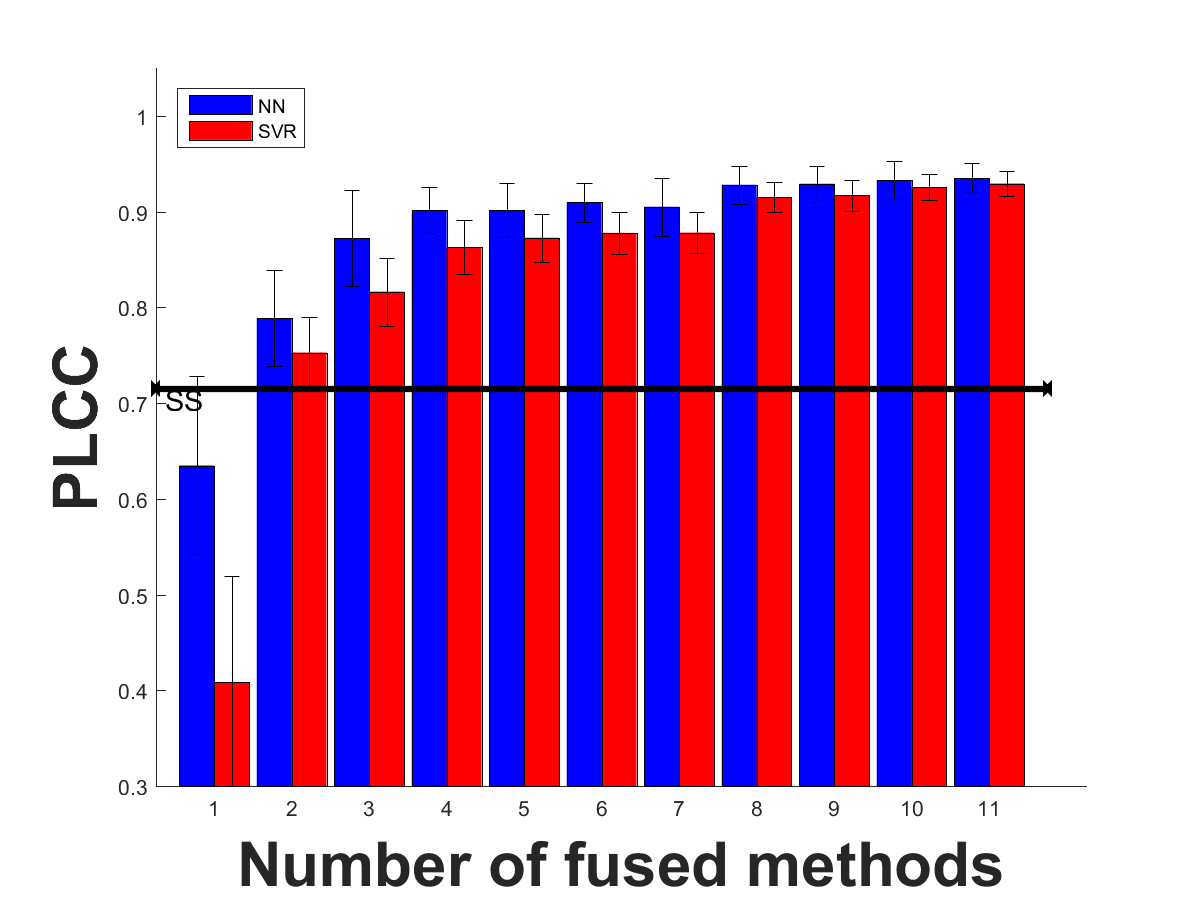}
  \vspace{0.03cm}
  \centerline{\footnotesize{(e) MULTI-PLCC }}

\end{minipage}
\hfill
\begin{minipage}[b]{0.32\linewidth}
  \centering
\includegraphics[width=\linewidth]{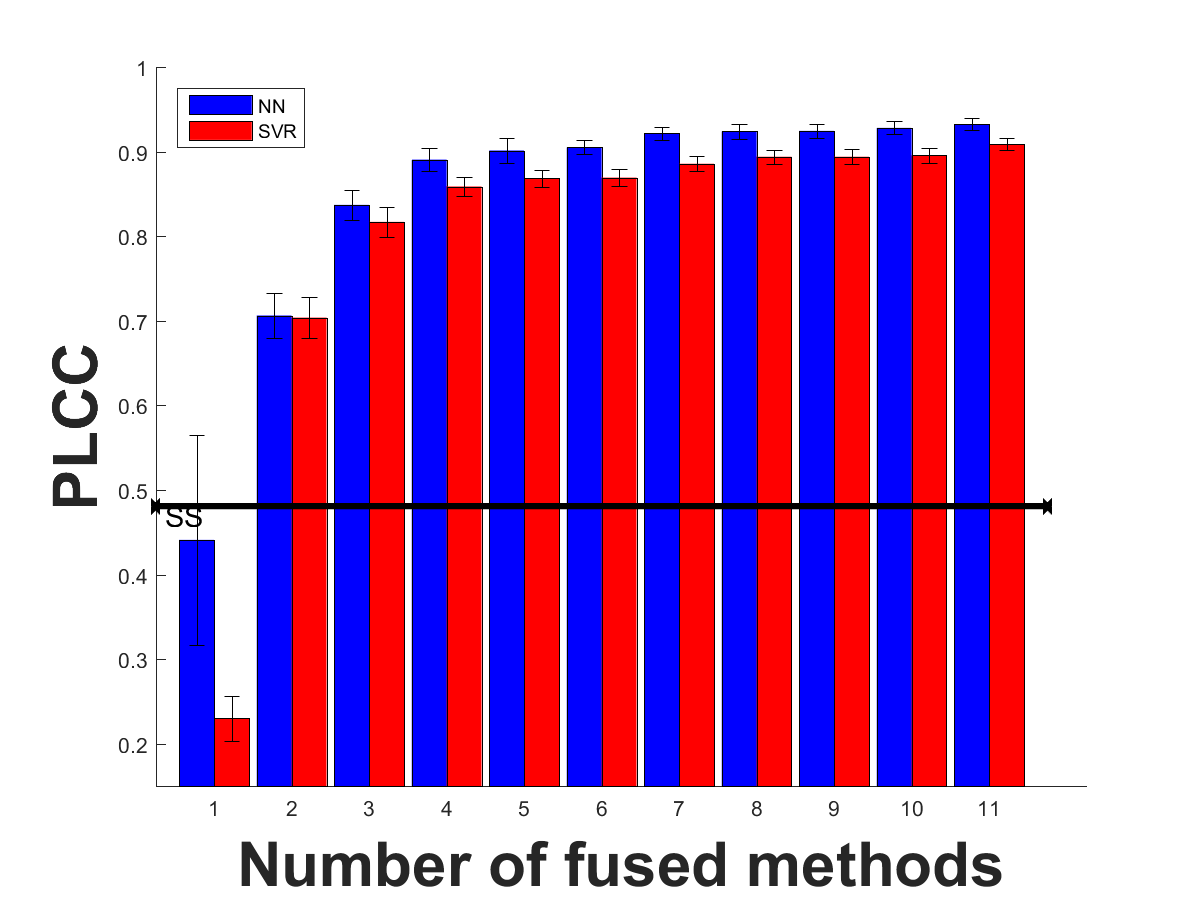}
  \vspace{0.03 cm}
  \centerline{\footnotesize{(f) TID-PLCC   } }

\end{minipage}

\begin{minipage}[b]{0.32\linewidth}
  \centering
\includegraphics[width=\linewidth]{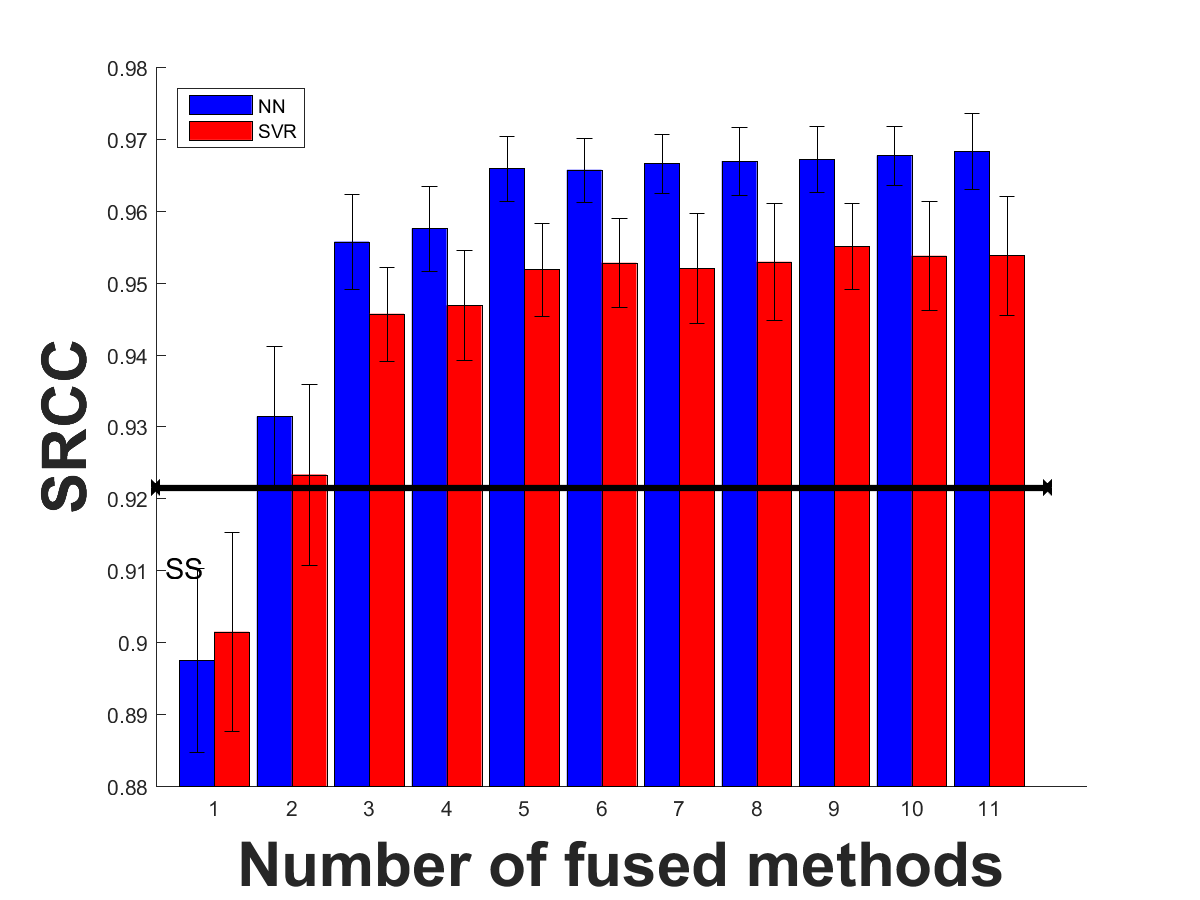}
  \vspace{0.03cm}
  \centerline{\footnotesize{(g) LIVE-SRCC}}

\end{minipage}
\hfill
\begin{minipage}[b]{0.32\linewidth}
  \centering
\includegraphics[width=\linewidth]{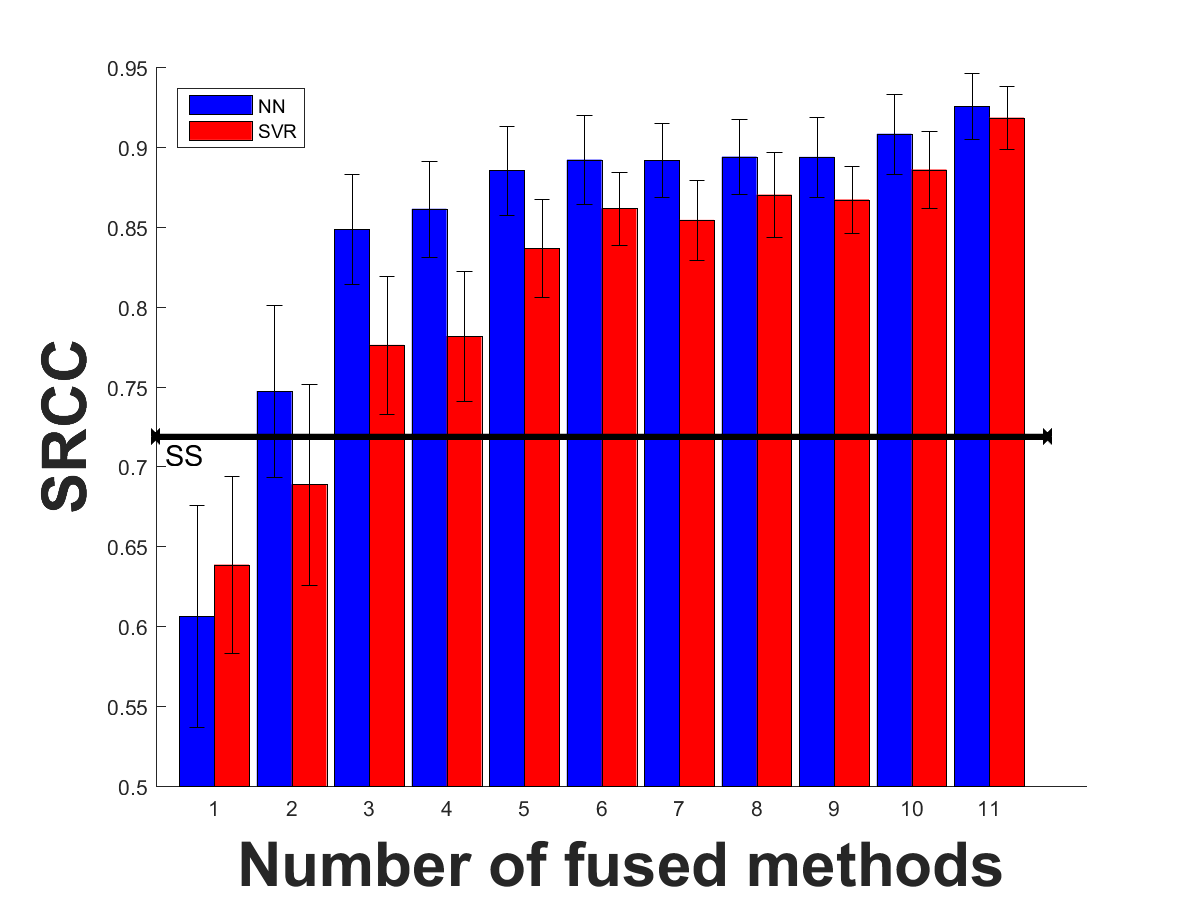}
  \vspace{0.03cm}
  \centerline{\footnotesize{(h) MULTI-SRCC }}

\end{minipage}
\hfill
\begin{minipage}[b]{0.32\linewidth}
  \centering
\includegraphics[width=\linewidth]{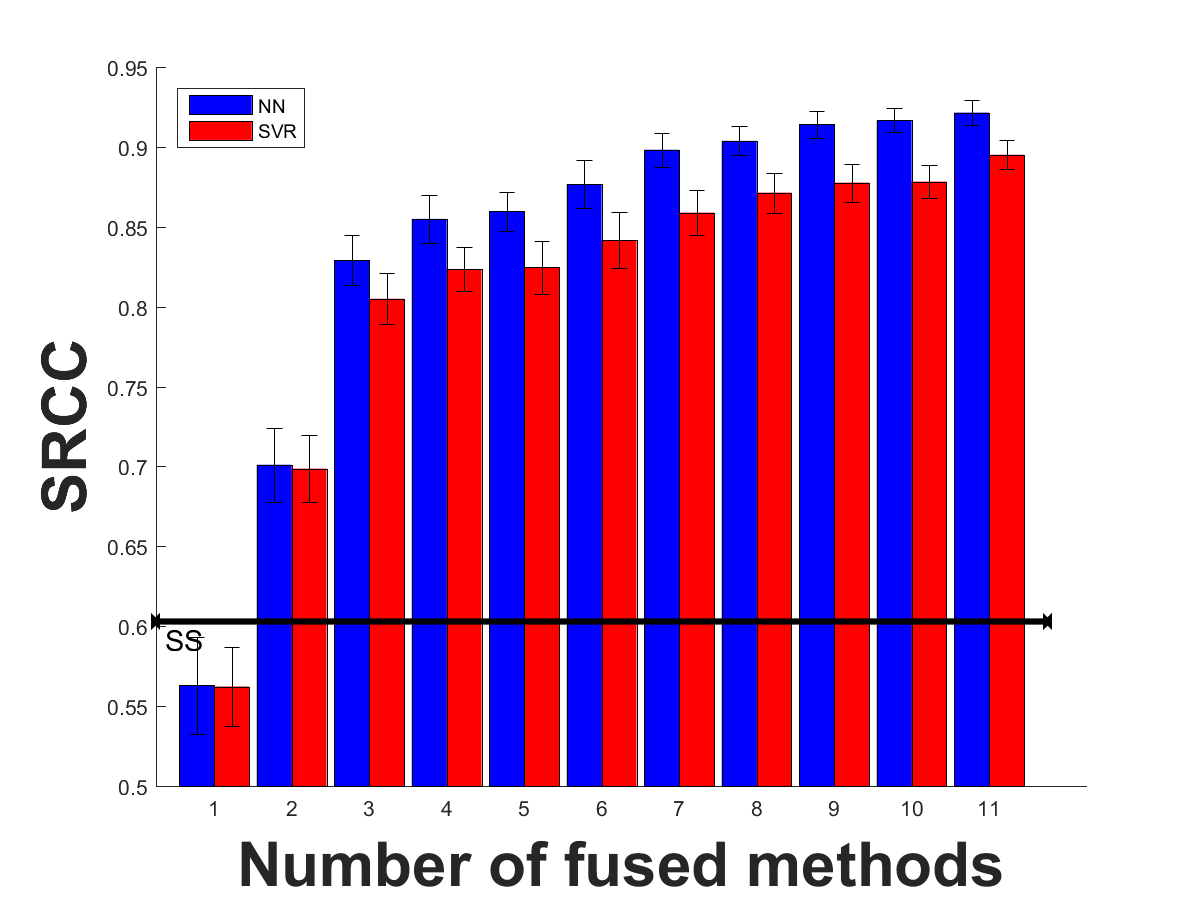}
  \vspace{0.03 cm}
  \centerline{\footnotesize{(i) TID-SRCC   } }

\end{minipage}
\caption{Performance of boosting methods versus number of fused methods.}
\label{fig:bar_both}

\end{figure*}

\end{center}

In  this section, we discuss the relative performance change as a consequence of adding new methods into the boosting algorithms. We start with the worst performing methods in each category and add the next best into the boosting in the next step. Based on the results in Table \ref{tab_part1}, we rank the methods for each database in a descending order in the root mean square error category, and in an ascending order in Pearson and Spearman correlation categories. The results are given in Fig. \ref{fig:bar_both} in which the lengths of the main bars correspond to the mean values and  the lengths of the thin bars plotted over the main bars show the standard deviations. We plot a horizontal black line in correlation figures, after which the increase in correlation coefficients becomes statistically significant with respect to the regressed worst performing quality estimator. Red bars correspond to the performance of support vector machine-based boosting and blue bars correspond to the neural network-based boosting.

As the number of fused methods increase, there is a general decrease in terms of root mean square error and an increase in terms of Pearson and Spearman correlations. Neural network-based boosting outperforms support vector machine-based boosting in terms of root mean square error in all the boosting scenarios when two or more methods are fused. Both Pearson and Spearman follow a non-decreasing behavior with respect to the number of fused methods other than a few exceptions. In terms of Pearson correlation, neural network-based boosting outperforms support vector machine-based boosting  in all the boosting scenarios. In terms of Spearman correlation, the worst performing quality estimators regressed with support vector machines perform slightly better than neural-network-based ones in the LIVE and the MULTI databases. However, in all the other scenarios, neural network-based boosting outperforms support vector machine-based boosting. 

\vspace{4.00mm}

\section{Conclusion}
\label{sec:conc}
We analyze the effect of boosting in image quality assessment using multi-method fusion. Experimental results show that boosting-based methods outperform existing best performing methods in $17$ out of $18$ comparisons and neural network-based boosting outperforms support vector machine-based boosting when two or more methods are fused. Based on these observations, we can claim that boosting generally enhances the performance of image quality assessment algorithms and the enhancement level depends on the type of the boosting strategy. Moreover, boosting the worst performing quality estimator with two or more additional methods leads to statistically significant improvements in all the scenarios independent of the boosting technique.

\vspace{-3.00mm}


\end{document}